\documentclass{pspum-l}

\newtheorem{theorem}{Theorem}[section]

\theoremstyle{definition}

\theoremstyle{remark}

\numberwithin{equation}{section}

\usepackage{graphicx,amsfonts,amsbsy, amssymb, amsthm, amsmath, mathrsfs}

\DeclareMathOperator{\tr}{tr}

\newcommand{\V}{\mathcal{V}}
\newcommand{\E}{\mathcal{E}}
\newcommand{\D}{\mathcal{D}}
\newcommand{\RH}{\mathcal{R}}
\newcommand{\Pau}{\mathcal{P}}

\newcommand{\U}{\mathrm{U}}
\newcommand{\UX}{\mathrm{X}}
\newcommand{\ui}{\mathrm{i}}
\newcommand{\ue}{\mathrm{e}}
\newcommand{\UI}{\mathrm{I}}
\newcommand{\J}{\mathrm{J}}
\newcommand{\ud}{\mathrm{d}}
\newcommand{\la}{\langle}
\newcommand{\ra}{\rangle}
\newcommand{\f}{\mathbf{f}}

\newcommand{\SU}{\mathrm{SU}}

\newcommand{\GSE}{\mathrm{GSE}}
\newcommand{\GOE}{\mathrm{GOE}}

\newcommand{\A}{{\mathbb A}}
\newcommand{\B}{{\mathbb B}}
\newcommand{\SM}{{\mathbb S}}

\newcommand{\ve}{\mathbf{e}}
\newcommand{\vsig}{\boldsymbol{\sigma}}
\newcommand{\vn}{\mathbf{n}}
\newcommand{\vv}{\mathbf{v}}
\newcommand{\vB}{\mathbf{B}}
\newcommand{\vf}{\mathbf{f}}

\newcommand{\vpsi}{\boldsymbol{\psi}}

\DeclareSymbolFont{lettersA}{U}{pxmia}{m}{it}

\DeclareMathSymbol{\alphaup}{\mathord}{lettersA}{"0B}
\DeclareMathSymbol{\betaup}{\mathord}{lettersA}{"0C}
\DeclareMathSymbol{\gammaup}{\mathord}{lettersA}{"0D}
\DeclareMathSymbol{\deltaup}{\mathord}{lettersA}{"0E}
\DeclareMathSymbol{\epsilonup}{\mathord}{lettersA}{"22}
\DeclareMathSymbol{\zetaup}{\mathord}{lettersA}{"10}
\DeclareMathSymbol{\etaup}{\mathord}{lettersA}{"11}
\DeclareMathSymbol{\thetaup}{\mathord}{lettersA}{"12}
\DeclareMathSymbol{\iotaup}{\mathord}{lettersA}{"13}
\DeclareMathSymbol{\kappaup}{\mathord}{lettersA}{"14}
\DeclareMathSymbol{\lambdaup}{\mathord}{lettersA}{"15}
\DeclareMathSymbol{\muup}{\mathord}{lettersA}{"16}
\DeclareMathSymbol{\nuup}{\mathord}{lettersA}{"17}
\DeclareMathSymbol{\xiup}{\mathord}{lettersA}{"18}
\DeclareMathSymbol{\piup}{\mathord}{lettersA}{"19}
\DeclareMathSymbol{\rhoup}{\mathord}{lettersA}{"1A}
\DeclareMathSymbol{\sigmaup}{\mathord}{lettersA}{"1B}
\DeclareMathSymbol{\tauup}{\mathord}{lettersA}{"1C}
\DeclareMathSymbol{\upsilonup}{\mathord}{lettersA}{"1D}
\DeclareMathSymbol{\phiup}{\mathord}{lettersA}{"1E}
\DeclareMathSymbol{\chiup}{\mathord}{lettersA}{"1F}
\DeclareMathSymbol{\psiup}{\mathord}{lettersA}{"20}
\DeclareMathSymbol{\omegaup}{\mathord}{lettersA}{"21}

\renewcommand{\Psi}{\varPsi}
\renewcommand{\Lambda}{\varLambda}
\renewcommand{\Sigma}{\varSigma}
\renewcommand{\Gamma}{\varGamma}
\renewcommand{\Theta}{\varTheta}
\renewcommand{\Xi}{\varXi}
\renewcommand{\Pi}{\varPi}
\renewcommand{\Upsilon}{\varUpsilon}
\renewcommand{\Phi}{\varPhi}

\newcommand{\R}{{\mathbb R}}

\newcommand{\Z}{{\mathbb Z}}
\newcommand{\T}{{\mathbb T}}
\newcommand{\C}{{\mathbb C}}

\newcommand{\coloneq}{\mathbin{\hbox{\raise0.08ex\hbox{\rm :}}\!\!=}}
\newcommand{\eqcolon}{\mathbin{=\!\!\hbox{\raise0.08ex\hbox{\rm :}}}}

\renewcommand{\epsilon}{\varepsilon}

\newcommand{\dpdot}{{\lower0.33ex\hbox{\LARGE$\cdot$}}}

\begin{document}


\title{Quantum graphs with spin Hamiltonians}

\author{J.M.~Harrison}
\address{Department of Mathematics, Baylor University,
Waco, TX 76798}
\email{jon\_harrison@baylor.edu}
\thanks{The author was supported in part by NSF Grants DMS-0604859 and DMS-0648786.}

\subjclass{Primary 34B45; Secondary 81Q10, 81Q50}
\date{October 30, 2007 and, in revised form, January 14, 2007.}


\keywords{Quantum graphs, Dirac operator, Rashba Hamiltonian}

\begin{abstract}
The article surveys quantization schemes for metric graphs with spin.
Typically quantum graphs are defined with the Laplace or Schr\"odinger operator which describe particles whose
intrinsic angular momentum (spin) is zero.  However, in many applications,
for example modeling an electron (which has spin-$1/2$) on a network of thin
wires, it is necessary to consider operators which allow spin-orbit
interaction.  The article presents a review of quantization schemes for graphs
with three such Hamiltonian operators, the Dirac, Pauli and Rashba
Hamiltonians.  Comparing results for the trace formula, spectral statistics and spin-orbit localization on quantum graphs with spin Hamiltonians.
\end{abstract}

\maketitle

\section{Introduction}\label{sec:int}

Quantum graphs have become a widely applied model in mathematical physics.  They fill a void between the simple quantum mechanics of the interval and more complex problems where it is often desirable to develop intuition in this quasi-one-dimensional system.  Motivation for this approach comes from the picture of quantum mechanics on a three dimensional network of thin wires.  It is expected that properties of the quantum graph will also be manifest in truly three dimensional systems, a subject of current research see e.g. \cite{exn:cos,gri:sgn,kuc:cos}.  Most work on quantum graphs has concentrated on the properties of the Laplace and Schr\"odinger operators, see \cite{kuc:qg1} for a review.  However, experiments on mesoscopic systems, systems with lengths on the scale of nanometers, measure features of electrons confined to these narrow wires.  Indeed quantum graphs were first proposed as models of free p-electrons in organic molecules like naphthalene, see \cite{kuc:gmw}.  The electron is a fermion possessing an additional quantum mechanical degree of freedom, intrinsic angular momentum called spin.  The effect of spin in the particles physics is highly nontrivial -- the connection between spin and particle statistics requires fermion eigenfunctions to be antisymmetric under particle exchange generating a wealth of phenomena in physics and chemistry.  For single particle quantum mechanics the effect of coupling between the spin dynamics and particle dynamics -- \emph{spin orbit coupling} -- also generates new physics observable in experiments.  Understanding how the presence of spin influences the quantum mechanics of graphs is therefore an important current question.

In this article I review the present state of knowledge of quantum mechanics on graphs where the Hamiltonian operator acts on multi-component wave functions that describe particles with spin.  The results fall into two broad categories, stemming from problems in quantum chaos and mesoscopic models of localization.
A good introduction to the application of graph models in quantum chaology is found in \cite{gnu:qga}.
One characteristic quantum phenomena associated with chaotic classical dynamics is  an energy level spectrum with statistics that resemble those of an ensemble of random matrices, this correspondence is known as the Bohigas-Gianoni-Schmidt conjecture \cite{boh:ccq}. The choice of random matrix ensemble depends on the symmetries of the quantum system.  For systems with half-integer spin time-reversal symmetry, in particular, takes a different form to that found in systems with integer spin.  This results in a doubly degenerate spectrum, \emph{Kramers' degeneracy}, and increased level repulsion between distinct eigenvalues.  A notable success in the application of quantum graphs with spin to quantum chaos has been to explain the effect of this change in symmetry in the spectral statistics of graphs with randomly chosen spin transformations at the vertices in terms of periodic orbits \cite{bol:sff,bol:scf}.  These arguments have recently been extended to transport properties of chaotic cavities with spin-orbit interaction \cite{bol:?}.

The Rashba Hamiltonian was introduced to model the two-dimensional electron gas confined on the surface of a narrow gap semiconductor \cite{ras:pos}.  The effect of the Rashba spin-orbit interaction on a graph can be detected in localization phenomena.  Numerical results for one-dimensional diamond chains show ideal localization, zero conductance, in the absence of a magnetic field \cite{ber:rei,ber:req}.  For the two-dimensional $\mathcal{T}_3$ lattice localization is not perfect but the spin-orbit coupling and magnetic field can be tuned to suppress the conductance.  Analytical results for the spectrum show Bloch bands that collapse to infinitely degenerate eigenvalues \cite{pan:lep}.

The article is organized as follows: Section \ref{sec:qg} introduces the nomenclature of quantum graphs.  In section \ref{sec:quantization} we collect results on the quantization of graphs with spin Hamiltonians.  Section \ref{sec:trace} discusses the trace formula for a quantum graph with spin  and in the balance we describe effects of spin-orbit interaction that have been investigated using quantum graphs in spectral statistics, section \ref{sec:ss}, and Rashba localization, section \ref{sec:localization}.

\section{Quantum graphs}\label{sec:qg}

I have endeavored to follow standard terminology for quantum and metric graphs throughout.  A reader already familiar with these ideas is encouraged to skipped ahead and refer back to this section if necessary.

A metric graph $G$ consists of a set $\V$ of vertices and a set $\E$ of edges. $N=|\E|$ denotes the number of edges.
Each edge $e=\{ v,w \}$ connects a pair of vertices $v,w\in \V$ and the edge $e$ is associated with an interval $[0,L_e]$ where $L_e\in [L_{\mathrm{min}}, L_{\mathrm{max}}]$ is the length of the edge,
$0<L_{\mathrm{min}}< L_{\mathrm{max}}<\infty$.
Introducing a coordinate $x_e \in [0,L_e]$ on the edge the graph naturally becomes directed with the initial vertex $v$ of the edge located at $x_e=0$ and the terminal vertex $w$ at $x_e=L_e$.  A directed edge will be denoted with parentheses, $e=(v,w)$.
Functions on the graph are defined by the set of functions on the edges $f=(f^1,\dots,f^{N})$.

The quantization of the graph requires the introduction of the Hilbert space,
\begin{equation}\label{eq:hilbert}
    L^2(G)=\bigoplus_{e\in \E} L^2(0,L_e) \ .
\end{equation}
Other spaces of functions are defined analogously, for example
$W^{2,2}(G)$ is the space of functions on $G$ with components drawn from the
$L^2$-Sobolev spaces $W^{2,2}(0,L_e)$.
 In this article we will be concerned with the properties of operators acting on multi-component functions on $G$ where the relevant Hilbert space is then $L^2(G)\otimes \C^n$ for spinor valued functions with $n$ components.
The scalar product of $f,g$ is defined by the scalar product on the edges,
\begin{equation}\label{eq:scalarproduct}
    \la f,g \ra = \sum_{e\in \E} \la f^e,g^e \ra_e
\end{equation}
where $\la f^e,g^e\ra_e$ is the scalar product in $L^2(0,L_e)\otimes \C^n$.

\section{Quantization schemes}\label{sec:quantization}

 The approach taken to quantizing a graph follows the same pattern for all the spin operators.  One takes the relevant operator in three dimensions, restricting it to one dimension to define the operator on an interval.  As spin is an intrinsic angular momentum naturally connected to rotations in three dimensions it is possible to restrict such three dimensional operators in different ways.  However, if the lengths of the edges can be chosen freely it is always possible to embed a graph in three dimensions in which case the three dimensional notion of a particles spin still makes sense.  This is also the natural physical situation to study where a quantum graphs is used to model networks of thin wires.  Having defined the operator on the edges matching conditions for the multi-component wave functions at the graph vertices are specified to make the operator self-adjoint.

\subsection{The Dirac operator}\label{sec:dirac}

The free Dirac operator on a graph was first considered by Bulla and Trenkler
\cite{bul:fdo}.  They take a two dimensional Dirac operator on the intervals, construct self-adjoint realizations and analyze certain special cases.
In \cite{bol:ssd} we introduced an approach which extends the classification of self-adjoint Laplace operators \cite{kos:krf} to classify self-adjoint realizations of the Dirac operator in terms of matching conditions for both two a four component spinor wavefunctions.
Further quantization schemes for a metric graphs with a Dirac operator have been developed by Post.  In \cite{pos:foa} he introduces the notion of a first order supersymmetric Dirac operator on both discrete and metric graphs in a very general framework and calculates the index of the operator proving that it agrees with the corresponding index of the discrete Dirac operator.  First order boundary triples are used to define a Dirac operator in \cite{pos:foo}.  In the following, for simplicity, we follow the approach we developed in \cite{bol:ssd}.

In one spatial dimension the Dirac equation is
\begin{equation}\label{eq:1ddirac}
\ui \hbar \frac{\partial}{\partial t} \Psi(x,t)
= \left( -\ui \hbar c \, \alpha \, \frac{\partial}{\partial x} +
mc^{2}\, \beta
 \, \right) \Psi(x,t) \ ,
\end{equation}
where $\alpha$ and $\beta$ are matrices that satisfy the Dirac algebra $\alpha^2=\beta^2=\UI$ and $\alpha\beta+\beta\alpha=0$.  It is natural to consider this equation on the line as deriving from a restriction of the Dirac operator in three spacial dimensions with the consequent notions of particles, antiparticles and spin all being carried over.  The equation then arises from implementing the Poincar\'e space-time symmetries in relativistic quantum mechanics, see \cite{b:tha:tde}, and the matrices $\alpha$ and $\beta$ are four dimensional, a convenient choice being
\begin{equation}\label{eq:3dirmat}
\alpha = \left( \begin{array}{cccc}
0 & 0 & 0 & -\ui \\
0 & 0 & \ui & 0 \\
0 & -\ui & 0 & 0 \\
\ui & 0 & 0 & 0 \\
\end{array} \right) \ ,
\qquad
\beta = \left( \begin{array}{cccc}
1 & 0 & 0 & 0 \\
0 & 1 & 0 & 0 \\
0 & 0 & -1 & 0 \\
0 & 0 & 0 & -1 \\
\end{array} \right).
\end{equation}
Alternatively one might look for the simplest faithful irreducible representation of the Dirac algebra which uses two dimensional matrices.  Here we consider only the first approach with four component spinor wave functions.  The quantization of two component wave functions was carried out following the same scheme in \cite{bol:ssd}.  We found that to make such a Dirac operator time-reversal symmetric (in a nontrivial way) it is necessary to replace each edge of the combinatorial graph with two directed bonds one running in each direction, effectively introducing four-components to the wave function on each edge.  The scattering matrices at the vertices and the spectrum of the two component operators on pairs of directed bonds are then equivalent to those of a four component Dirac operator.

A free Dirac operator on an edge has the form
\begin{equation}\label{eq:3dop}
\D_e
= -\ui \hbar c \, \alpha \, \frac{\ud}{\ud x_e} +
mc^{2}\, \beta \ .
\end{equation}
$\D^0$ is defined as $\D$ with the domain $W^{2,1}_0(G)\otimes \C^4$, the Sobolev space of functions $f$ on the graph where the components (and their first derivatives) are $L^2(0,L_e)$ and the functions vanish at graph vertices, $f^e(0)=f^e(L_e)=0$.
Self-adjoint realizations of $\D$ are defined by extensions of the closed symmetric operator $\D^0$ to domains that are subspaces of $W^{2,1}_0(G)\otimes \C^4$ isotropic with respect to a skew Hermitian form
$\Omega (f,g)=\la \D f,g\ra - \la f,\D g \ra$.

Integrating by parts one may rewrite $\Omega$ as a complex symplectic form on $\C^{8N}$ depending only on values of the spinors at the graph vertices.
\begin{equation}\label{eq:omegascalar}
  \Omega(f,g)= ( \begin{array}{cc}
  \f_{+}^{\dagger}& \f_{-}^{\dagger} \\
\end{array} ) \left( \begin{array}{cc}
0 & \UI_{4B} \\
-\UI_{4B} & 0 \\
\end{array} \right) \left( \begin{array}{c}
\boldsymbol{g}_{+}\\
\boldsymbol{g}_{-}\\
\end{array} \right) \ ,
\end{equation}
where
{\setlength\arraycolsep{2pt}
\begin{equation*}
\begin{split}
\f_{+}^{T}
&=\bigl( f_{1}^{1}(0),f_{2}^{1}(0),\dots,f_{1}^{N}(0),f_{2}^{N}(0),
f_{1}^{1}(L_{1}), f_{2}^{1}(L_{1}), \dots ,f_{1}^{N}(L_{N}),
f_{2}^{N}(L_{N}) \bigr) \ , \\
\f_{-}^{T}
&=\bigl( -f_{4}^{1}(0),f_{3}^{1}(0),\dots,-f_{4}^{N}(0),f_{3}^{N}(0),  f_{4}^{1}(L_{1}),-f_{3}^{1}(L_{1}), \dots ,f_{4}^{N}(L_{N}),-f_{3}^{N}(L_{N}) \bigr).
\end{split}
\end{equation*}
  The vector $\f_{+}$ contains the first and second components of the spinors and $\f_-$ the third and fourth components.  Maximal isotropic subspaces with respect to $\Omega$ correspond to maximal subspaces of vectors in $\C^{8N}$ on which the symplectic form (\ref{eq:omegascalar}) vanishes.

As in the case of the Laplace operator a linear subspace of vectors in $\C^{8N}$ will be defined as those vectors satisfying the equation
\begin{equation}\label{eq:matching}
\A \f_{+} + \B \f_{-} = 0  \ ,
\end{equation}
with complex matrices $\A$ and $\B$.  In order for the space to have dimension $4N$ the $4N\times 8N$ matrix $(\A,\B)$ must have maximal rank.  Kostrykin and Schrader \cite{kos:krf} showed that such a subspace is maximally isotropic if and only if $(\A,\B)$ has maximal rank and $\A\B^\dagger$ is Hermitian.

\begin{theorem}\label{self-adjoint Dirac}
$\A \f_{+} + \B \f_{-} = 0$ defines matching conditions of spinor wave functions at the vertices of a graph for a set of $N$ bonds.  These matching conditions yield a self-adjoint realization of the Dirac operator on the graph if and only if
 \begin{equation*}
    \mathrm{rank}(\A,\B)=4N \quad \textrm{and} \quad \A\B^\dagger=\B\A^\dagger \ .
 \end{equation*}
\end{theorem}

When considering graphs it is often convenient to work with scattering matrices at the vertices rather than with the matching conditions directly.  To obtain the vertex scattering matrix of the Dirac operator we note that eigenfunctions on an edge, the solutions of $\D \psi^e(x_e)=E \psi^e(x_e)$, are plane waves.
\begin{equation}\label{eq:5planewave}
\begin{split}
\psi_{k}^{e}(x_e)&=\mu_{\alpha}^{e} \left( \begin{array}{c}
1 \\
0 \\
0 \\
\ui \gamma (k) \\
\end{array} \right) \ue^{\ui k x_e} +
 \mu_{\beta}^{e} \left( \begin{array}{c}
0 \\
1 \\
-\ui  \gamma (k) \\
0 \\
\end{array} \right) \ue^{\ui k x_e}  \\ &\quad
+ \hat{\mu}_{\alpha}^{e} \left( \begin{array}{c}
1 \\
0 \\
0 \\
-\ui  \gamma (k)\\
\end{array} \right) \ue^{-\ui k x_e} +
\hat{\mu}_{\beta}^{e} \left( \begin{array}{c}
0 \\
1 \\
\ui  \gamma (k)\\
0 \\
\end{array} \right) \ue^{-\ui k x_e} \ ,
\end{split}
\end{equation}
for $k>0$ and positive energy $E$ and $\gamma(k)$ given by
\begin{equation}\label{eq:3dispersion}
\gamma(k) = \frac{E-mc^2}{\hbar c k} \ ,\qquad
E= \sqrt{ (\hbar c k)^{2} +m^{2}c^{4} } \ .
\end{equation}
Consider a vertex $v$ of valency $d_v$ whose matching conditions are defined by a pair of matrices $\A^{(v)},\B^{(v)}$, so $\A^{(v)} \vpsi_{+} + \B^{(v)} \vpsi_{-}=0$.  To simplify the formulae we take edges meeting at $v$ to be aligned with $v$ at $x_e=0$.
We may define two vectors of plane-wave coefficients for the incoming and outgoing
waves respectively,
\begin{equation}
\begin{split}
\boldsymbol{\overleftarrow{\mu}}&=\big(
\hat{\mu}^{1}_{\alpha},
\hat{\mu}^{1}_{\beta},
\dots,
\hat{\mu}^{d}_{\alpha},
\hat{\mu}^{d}_{\beta} \big)^T \ ,\\
\boldsymbol{\overrightarrow{\mu}}&=\big(
\mu^{1}_{\alpha},
\mu^{1}_{\beta},
\dots,
\mu^{d}_{\alpha},
\mu^{d}_{\beta} \big)^T \ .
\end{split}
\end{equation}
Then $\vpsi_{+}=\boldsymbol{\overrightarrow{\mu}}+\boldsymbol{\overleftarrow{\mu}}$ and  $\vpsi_{-}=-\ui \gamma(k) (\boldsymbol{\overrightarrow{\mu}} - \boldsymbol{\overleftarrow{\mu}})$.  Consequently,
\begin{equation}
\boldsymbol{\overrightarrow{\mu}}=-(\A^{(v)} - \ui \gamma (k) \B^{(v)})^{-1}
(\A^{(v)} + \ui \gamma (k) \B^{(v)})\,
\boldsymbol{\overleftarrow{\mu}} \ .
\end{equation}
The vertex transition matrix
\begin{equation}\label{eq:4transition}
 \T^{(v)}=-(\A^{(v)} - \ui \gamma (k) \B^{(v)})^{-1}
 (\A^{(v)} + \ui \gamma (k) \B^{(v)})
\end{equation}
relates coefficients of incoming and outgoing plane-waves at $v$.  It is unitary as $\A^{(v)}\B^{(v)\dagger}=\B^{(v)}\A^{(v)\dagger}$, which ensures current
conservation at the vertex.

\subsection{Time-reversal invariance for the Dirac operator}

Time-reversal symmetry in a system with half-integer spin takes a different form from the symmetry under complex conjugation familiar for scalar functions, see \cite{b:haa:qsc}.  The time-reversal operator $T$ for systems with half-integer spin remains anti-unitary but it squares to minus the identity, $T^2=-\UI$.  For any Hamiltonian $H$ invariant under time reversal $T$ commutes with $H$ and if $\psi$ is an eigenfunction so is $T\psi$.  However, when $T^2=-\UI$ the two eigenfunctions $\psi$ and $T\psi$ are orthogonal and eigenvalues of $H$ are doubly degenerate -- known as Kramers' degeneracy.

For the Dirac operator on a graph to be invariant under time-reversal the operator $T$ must commute with the Hamiltonian which forces the mass term to vanish fixing $\gamma(k)=1$.  For the vertex transition matrices we find time-reversal symmetry implies
\begin{equation}\label{eq:trs}
(\T^{(v)})^{T}= - \left( \UI_{d_v} \otimes \J \right)
\T^{(v)} \left( \UI_{d_v} \otimes \J \right) \ ,
\end{equation}
where
\begin{equation}
\J=
\begin{pmatrix}
0 & 1 \\
-1 & 0 \\
\end{pmatrix} \ .
\end{equation}
 If we consider the $2\times 2$ block of $\T^{(v)}$ that relates the pair of incoming spinors arriving from edge $e$ to the outgoing pair on $f$ (\ref{eq:trs}) is equivalent to,
\begin{equation}
(\T^{(v)})^{fe}= \bigl| (\T^{(v)})^{ef} \bigr| \,
\left( (\T^{(v)})^{ef} \right) ^{-1} \ .
\end{equation}
This suggests a simple method of constructing time-reversal symmetric
transition matrices that relates transition matrices of the Dirac operator to those of the Schr\"odinger operator on the graph.  Let
\begin{equation}\label{eq:genT}
\T^{(v)}= (\U^{(v)})^{-1} ( \UX^{(v)} \otimes \UI_2 ) \U^{(v)} \ ,
\end{equation}
where $\U^{(v)}$ is a block diagonal matrix $\U^{(v)}=\textrm{diag}\{ u_1,\dots,u_{d_v} \}$ with $u_j \in  \SU(2)$ and $\UX$ is a unitary symmetric $d_v \times d_v$ matrix.  Then $\T^{(v)}$ respects (\ref{eq:trs}) and results from a self-adjoint realization of the Dirac operator on $G$.
$\UX$ can be thought of as the transition matrix of the graph with out spin.  For the Laplace operator on the graph time-reversal invariant matching conditions at the vertices generate a symmetric unitary vertex transition matrix.
The elements $u_j\in \SU(2)$ generate spin rotations at $v$.
We may now define, for example, Neumann like matching conditions for the Dirac operator at the vertex $v$.
\begin{equation}\label{eq:Dirac Neumann}
\begin{split}
    (u^{(v)}_{e_i})^{-1} \vf_+^{e_i} =
    (u^{(v)}_{e_j})^{-1} \vf_+^{e_j}
    &=\vf_+(v) \qquad  e_i,e_j \sim v \\
    \sum_{e\sim v} (u^{(v)}_{e})^{-1} \vf_-^{e} &= 0
    \end{split}
\end{equation}
The sum is over edges $e$ connected to $v$, $e\sim v$.
 These conditions can be represented in the form (\ref{eq:matching}) by matrices
\begin{equation}
\A^{(v)}=\left\{ \left( \begin{array}{ccccc}
1 & -1 & 0 & 0 & \dots \\
0 & 1 & -1 & 0 & \dots \\
& & \ddots & \ddots & \\
0 & \dots & 0 & 1 & -1 \\
0  & \dots & 0& 0 & 0 \\
\end{array} \right) \otimes \UI_2 \right\}\U^{(v)} \quad
\B^{(v)}= \left\{ \left( \begin{array}{cccc}
0 & 0 &\dots & 0 \\
\vdots & \vdots & & \vdots \\
0 & 0 &\dots & 0 \\
1 & 1 &\dots & 1 \\
\end{array} \right) \otimes \UI_2 \right\}\U^{(v)}
\end{equation}
Substituting the matching conditions in (\ref{eq:4transition}) produces a transition matrix of the form (\ref{eq:genT}) with $[\UX^{(v)}]_{ef}=2/v_d - \delta_{ef}$ the familiar transmission amplitudes of the Laplace operator on a graph with Neumann matching conditions at the vertices, see \cite{kot:pot}.


\subsection{The Pauli operator}\label{sec:pauli}

For comparison we introduce a quantization scheme for a simple form of the Pauli operator which is closely related to the Laplace quantization \cite{kos:krf}.
Consider the following operator on the edges of the graph,
\begin{equation}\label{eq:pauli op}
   \Pau_e = -\frac{\ud}{\ud x_e^2}+\vB_e . \vsig
\end{equation}
where $\vsig$ is the vector of Pauli matrices and $\vB_e$ is a constant vector depending on the edge.  The operator acts on two component functions $f_e$ in the Hilbert space $L^2(0,L_e)\otimes \C^2$.

$\Pau^0$ is defined as $\Pau$ with the domain $W^{2,2}_0(G)\otimes \C^2$, the Sobolev space of functions $f$ on the graph where the components (and their first and second derivatives) are in $L^2(0,L_e)$ and $f_e(0)=f_e(L_e)=f_e'(0)=f_e'(L_e)=0$.
Self-adjoint realizations of $\Pau$ are again defined by extensions of the closed symmetric operator $\Pau^0$ to domains that are subspaces of $W^{2,2}(G)\otimes \C^2$ isotropic with respect to a skew Hermitian form
$\Omega (f,g)=\la \Pau f,g\ra - \la f,\Pau g \ra$.
As $(\vB_e.\vsig)^\dag= \vB_e.\vsig$ the form $\Omega (f,g)$ is the same as that obtained for the Laplace operator where the number of edges is doubled.
A domain on which the operator is self-adjoint is specified by a pair of
$4N\times 4N$ matrices $\A$ and $\B$, see \cite{kos:krf}, where $(\A,\B)$ has maximal rank and $\A\B^\dagger$ is Hermitian via the linear equation,
\begin{equation}\label{eq:paulibc}
    \A \vf +\B \vf' = 0 \ .
\end{equation}
$\vf$ and $\vf'$ are vectors of the components of $f$ and its derivative evaluated at the ends of the intervals.
\begin{equation*}\label{eq:defnvf}
\begin{split}
    \vf &=\big( f_1^1 (0), f_2^1 (0) \dots, f_1^N(0), f_2^N(0),
    f_1^1(L_e), f_2^1(L_e),\dots, f_1^N(L_e), f_2^N(L_e) \big)^T \\
    \vf' &=\big( f_1^{\prime 1} (0), f_2^{\prime 1}  (0) \dots, f_1^{\prime N}(0), f_2^{\prime N}(0),
    -f_1^{\prime 1}(L_e), f_2^{\prime 1}(L_e),\dots, -f_1^{\prime N}(L_e), f_2^{\prime N}(L_e) \big)^T
\end{split}
\end{equation*}

To define the vertex transmission matrix it is convenient to diagonalize the spin transformation on the edge,
\begin{equation}\label{eq:diag B.sig}
    \vB_e .\vsig = |\vB_e| u_e \, \sigma_3 \, u_e^{-1} \ ,
\end{equation}
where $u_e$ is in $\SU(2)$.  Then plane wave solutions for the eigenproblem
$\Pau_e \psi_e = \lambda \psi_e$ are given by
\begin{equation}\label{eq:pauli planewaves}
    f_e=u_e\begin{pmatrix}
    \mu^e_1 \ue^{\ui k^e_1} +\hat{\mu}^e_1 \ue^{-\ui k^e_1}\\
    \mu^e_2 \ue^{\ui k^e_2} +\hat{\mu}^e_2 \ue^{-\ui k^e_2}\\
    \end{pmatrix} \ ,
\end{equation}
$(k^e_2)^2=(k^e_1)^2+2|\vB_e|$ and $\lambda=(k^e_1)^2+|\vB_e|$.
To obtain the vertex scattering matrix consider a vertex $v$ of valency $d_v$ with the edges aligned so $v$ is the initial vertex of each edge.
Vectors of incoming and out going plane-wave coefficients are respectively,
\begin{equation}
\begin{split}
\boldsymbol{\overleftarrow{\mu}}&=\big(
\hat{\mu}^{1}_{1},
\hat{\mu}^{1}_{2},
\dots,
\hat{\mu}^{d}_{1},
\hat{\mu}^{d}_{2} \big)^T \ ,\\
\boldsymbol{\overrightarrow{\mu}}&=\big(
\mu^{1}_{1},
\mu^{1}_{2},
\dots,
\mu^{d}_{1},
\mu^{d}_{2} \big)^T \ .
\end{split}
\end{equation}
Then defining $K^{(v)}=\textrm{diag}\{k_1^1,k_2^1,k_1^2,\dots,k_2^{d_v}\}$
we have
\begin{equation}\label{eq:in}
\begin{split}
  \vpsi & =U^{(v)}( \boldsymbol{\overrightarrow{\mu}}+\boldsymbol{\overleftarrow{\mu}}) \ ,\\
  \vpsi'& = -\ui U^{(v)} K^{(v)}  (\boldsymbol{\overrightarrow{\mu}} - \boldsymbol{\overleftarrow{\mu}}) \ ,
\end{split}
\end{equation}
where $U^{(v)}$ is again a block diagonal matrix,
$U^{(v)}=\textrm{diag}\{u_1,\dots,u_{d_v} \}$.
Consequently substituting in (\ref{eq:paulibc}) the vertex scattering matrix
is
\begin{equation}\label{eq:4transition2}
 \T^{(v)}=- \big( U^{(v)} \big)^{-1} \left( \A^{(v)} - \ui K^{(v)} \B^{(v)}\right)^{-1}
 \left(\A^{(v)} + \ui K^{(v)} \B^{(v)}\right) U^{(v)}
 \ .
\end{equation}
In this form the transition matrix has built in spin rotations at the vertices of a similar form to that described in (\ref{eq:genT}) for the Dirac operator.
 A Pauli operator can also be defined for systems with higher spins, acting on multi-component wave functions of higher dimensions.  This was used in \cite{bol:sff} to extend results on the spectral statistics of the Dirac operator to any spin.

\subsection{The Rashba Hamiltonian}\label{sec:rashba}

The Rashba Hamiltonian, introduced in \cite{ras:pos} to model electronic structure of confined narrow-gap semiconductors, has also been studied as an operator on quantum graphs.  Harmer \cite{har:sfr} constructs the scattering matrix of the Rashba Hamiltonian on a ring with an arbitrary number of semi-infinite wires attached to investigate spin filtering.  In section \ref{sec:localization} we will discuss localization phenomena generated by Rashba spin-orbit coupling \cite{ber:rei,ber:req,pan:lep}.  So to complete our survey of quantization schemes we now turn to the Rashba Hamiltonian on a graph.

With the Rashba Hamiltonian it is useful to have in mind a graph with straight edges embedded in the $x$-$y$ plane in $\R^3$.  We will denote by $\ve_{(vw)}=(e_{(vw)x},e_{(vw)y},0)$ a unit vector in the direction $(vw)$.  Let
$A_{(vw)}$ denote the magnetic potential on $(vw)$, the component of a magnetic vector potential in the direction $\ve_{(vw)}$.  For simplicity we take $A_{(vw)}$ to be constant, as is the case with a uniform magnetic field $\vB$ where $A_{(vw)}=\frac{1}{2} (\vB\times \vv).\ve_{(vw)}$ and $\vv$ is a vector locating the vertex $v$.
$k_R$ is the Rashba coupling constant and $\vn=(0,0,1)$ is a unit vector perpendicular to the $x$-$y$ plane.  The Rashba Hamiltonian on an edge $e$ reads,
\begin{equation}\label{eq:rashba hamiltonian}
\begin{split}
   \RH_e&= \left( -\ui \frac{\ud}{\ud x_e} - A_e \right)^2 + 2 k_R
   \left(-\ui \frac{\ud}{\ud x_e} - A_e \right) (\vsig \times \vn ) . \ve \ , \\
   &=\left( \ui \frac{\ud}{\ud x_e} + A_e +k_R \sigma_e \right)^2 -k_R^2 \ .
   \end{split}
\end{equation}
$\vsig$ denotes the vector of $2\times 2$ Pauli matrices and
\begin{equation}\label{eq:sigvw}
    \sigma_e=\left( \begin{array}{cc}
    0 & e_y + \ui e_x\\
    e_y-\ui e_x& 0\\
    \end{array}
    \right) \ .
\end{equation}
A self-adjoint realization of the operator with delta type matching conditions at the vertices was defined by Pankrashkin in \cite{pan:lep}.
The matching conditions at the vertices are analogous to delta type conditions of the Schr\"odinger operator where $\epsilon(v)=0$ defines  Neumann like coupling at the vertex $v$.

\begin{theorem}\label{thm:sa Rashba}
Denote by an operator $\RH$ the Rashba Hamiltonian on the edges of the graph
acting on functions $f\in W^{2,2}(G)\otimes\C^2$ satisfying at each vertex $v\in \V$,
\begin{equation*}
    \vf_{(vw)}(0)=\vf_{(uv)}(L_{(uv)})=\vf(v), \quad (vw),(uv)\in \E \ ,
\end{equation*}
\begin{equation*}
\begin{split}
    \sum_{(uv)\in \E}& \left(\frac{\ud}{\ud x_{(uv)}} -\ui (A_{(uv)} +k_R \sigma_{(uv)})
    \right) \vf_{(uv)}(0)\\ &
    -\sum_{(vw)\in \E}
    \left(\frac{\ud}{\ud x_{(vw)}} -\ui (A_{(vw)} +k_R \sigma_{(vw)})
    \right) \vf_{(vw)}(L_{(vw)})
    = \epsilon(v)\vf(v) \ .
    \end{split}
\end{equation*}
Then $\RH$ is self-adjoint.
\end{theorem}

\section{The trace formula}\label{sec:trace}

A trace formula for the Laplacian on a graph was first produced by Roth
\cite{rot:lsl}.
Here the trace formula for the density of states of the free Dirac and Pauli operators is stated in the form developed by Kottos and Smilansky \cite{kot:qcg,kot:pot} for the Schr\"odinger operator.  A full review of the state of the art for the trace formula of a Schr\"odinger operator on a graph can be found in this volume \cite{bol:tv}.

The spectrum of a quantum graph can be computed using the scattering matrix of the graph.
The scattering matrix is a unitary $4N\times 4N$ quantum evolution operator acting on the vector of all plane-wave coefficients.  It is defined via
\begin{equation}\label{eq:3S}
\SM^{(uv)(wx)}(k)= \delta_{vw} \ \sigma_{(uv)(vx)} u_{(uv)(vx)} \ \ue^{\ui kL_{(vx)}} \ ,
\end{equation}
where $\sigma_{(uv)(vx)}=\big|(\T^{(v)})^{(uv)(vx)}\big|$ and
$u_{(uv)(vx)}=\big(\T^{(v)}\big)^{(uv)(vx)}/\big|(\T^{(v)})^{(uv)(vx)}\big|$.
The $k$-spectrum of the Dirac operator are the solutions of the
secular equation,
\begin{equation}\label{eq:secular}
    |\UI - \SM(k)|=0 \ .
\end{equation}
This quantization condition is obtained by considering vectors of plane-wave coefficients invariant under multiplication by $\SM(k)$ which define eigenfunctions on the graph.

The density of states is a comb of delta functions located at wave numbers $k_n$, $d(k) = \sum_{n=1}^\infty \delta(k-k_n)$.  The trace formula relates
this spectral distribution to the set of periodic orbits on the graph.  A periodic orbit $p$ of $n$ steps is a sequence $p=(e_1,e_2,\dots,e_n)$ of directed edges on the graph, where the terminal vertex of $e_j$ is the initial vertex of $e_{j+1}$ and $e_n$ terminates at the origin of $e_1$.
We will denote the set of Periodic orbits of $n$ bonds $P_n$ and
and $l_{p}$ is the metric length of the orbit $p$.  The total length of the
graph is $L=\sum_{e\in\E} L_{e}$.  A periodic orbit $p$ may be a repartition of a shorter primitive orbit in which case $r_p$ denotes the number of repartitions of the primitive orbit used to produce $p$.
\begin{theorem}
For the Dirac operator on a graph the density of states is,
\begin{equation*}\label{eq:traceD}
d(k)=\frac{2L}{\pi}+ \frac{1}{\pi} \sum_{p} \frac{l_{p}}{r_{p}}
A_{p} \, \tr(d_{p}) \cos (k l_{p}) \ ,
\end{equation*}
where
\begin{equation*}
\begin{split}
A_{p}&= \sigma_{e_1 e_2} \sigma_{e_2 e_3} \dots \sigma_{e_n e_1} \ ,\\
d_{p}&=u^{e_1 e_2} u^{e_2 e_3} \dots u^{e_n e_1} \ .
\end{split}
\end{equation*}
\end{theorem}
The trace formula has the same form as that of the Schr\"odinger operator on a graph.  The first term is a Weyl term, the mean density of states. The second contribution, the oscillating part, contains information about correlations in the spectrum written as a sum over all periodic orbits.
The difference from the scalar version appears in the additional trace of $d_p \in \SU (2)$ which is the product of the spin transformations at the vertices picked up while traversing the orbit $p$.  A similar term appears in the semiclassical Gutzwiller type trace formula for the Dirac operator \cite{bol:ste,bol:sad}.
The trace formula is derived from the secular equation (\ref{eq:secular}) in \cite{bol:ssd}.
For the Pauli operator on the graph a similar formula holds and the elements $u^{b_j b_{j+1}} \in \SU(2)$ may be replaced by a higher dimensional irreducible representation, see \cite{bol:sff}.

\section{Spectral statistics}\label{sec:ss}

Quantum graphs with spin Hamiltonians provide a simple model in which to investigate the effects of spin-orbit interaction.  Our first example is the effect of spin-orbit interaction on the statistics of the graph spectrum.

For systems with half-integer spin and time-reversal symmetry eigenvalues of the Hamiltonian are doubly degenerate.  When discussing spectral statistics it is therefore convenient to lift Kramers' degeneracy and also rescale the spectrum so that the mean spacing is one, for more details see \cite{b:haa:qsc}.  In the following we will assume such an unfolding of the spectrum has already taken place.

The spectrum of the Dirac operator can be investigated numerically using the secular equation (\ref{eq:secular}).  A readily available statistic for computation is the nearest neighbor spacing distribution $p(s)$, the probability density of the spacing $s$ between consecutive levels.  Figure \ref{fig:fcs} shows a histogram of the nearest neighbor spacing distribution and the integrated distribution for the Dirac operator on a fully connected square.
The spin transformations at the vertices were chosen randomly from $\SU(2)$.
The numerical results show good agreement with the spectral statistics of the Gaussian symplectic ensemble (GSE) of random matrices -- the ensemble of Hermitian matrices with probability measure invariant under symplectic transformations.  This correspondence with random matrix statistics, seen in a wide variety of chaotic quantum systems, is known as the Bohigas-Giannoni-Schmidt conjecture \cite{boh:ccq}.  The point of particular interest to us here is the ensemble of random matrices to which the statistics of the quantum spectrum correspond.  For systems with time-reversal symmetry and half-integer spin like the Dirac operator on a graph one generically sees GSE statistics, while for integer spin the corresponding random matrix statistics are those of the Gaussian orthogonal ensemble (GOE).

\begin{figure}[tb]
\begin{center}
\includegraphics[width=10cm]{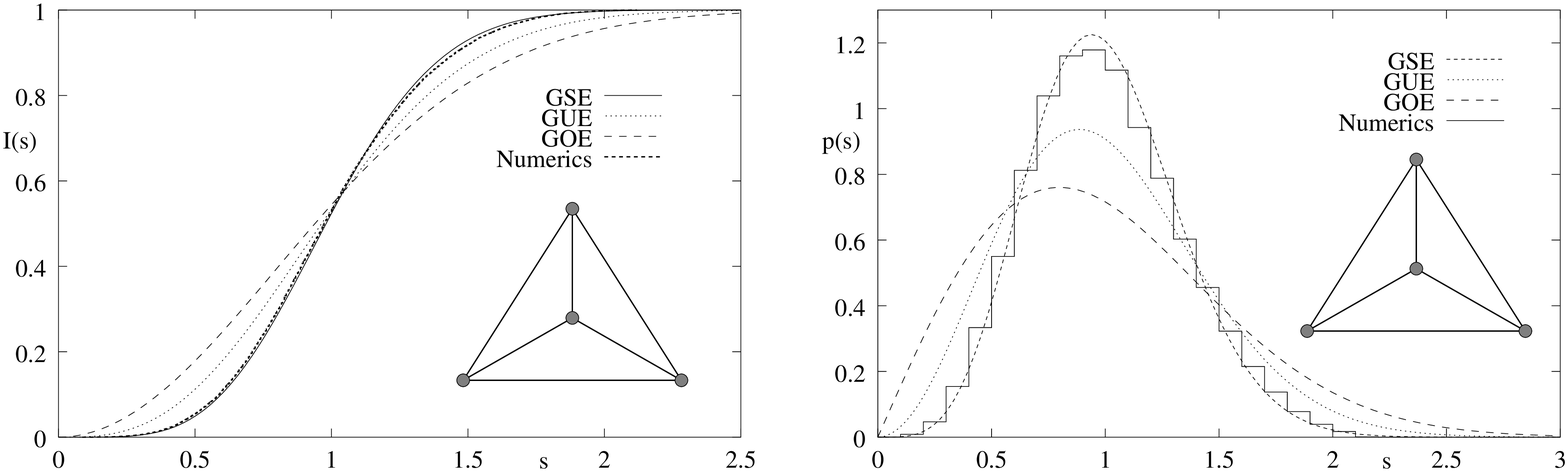}
\end{center}
\caption{The integrated nearest neighbor spacing distribution and
a histogram of the spacing distribution for a Dirac operator with Neumann
matching conditions on a fully
connected graph with four vertices, 24234 levels.  Figure taken from \cite{bol:ssd}.}
\label{fig:fcs}
\end{figure}

The two point correlation function of the unfolded spectrum is,
\begin{equation}
R_{2}(x) =  \lim_{\Lambda \rightarrow
\infty} \frac{1}{\Lambda} \int_{0}^{\Lambda} d(y)\,d (y- x)
\ud y -1 \ .
\end{equation}
where $d(y)$ is the unfolded density of states.
The fourier transform of $R_2$, the form factor, has been the spectral statistic most amenable to
theoretical investigation.
Using the trace formula for the Dirac operator (\ref{eq:traceD}) the spectral form factor
can be written as a sum over pairs of periodic orbits,
\begin{equation}\label{eq:KDirac}
K_{\textrm{Dirac}}(\tau) = \frac{1}{4(2L)^2}
\sum_{p,q} \frac{l_{p}l_{q}}{r_{p}r_{q}} A_{p} A_{q} \,
\ue^{\ui\pi(\mu_p +\mu_q)} \tr (d_{p}) \tr (d_{q}) \,
\delta \left( \tau  - \frac{l_{q}}{2L} \right) \delta_{l_{p},l_{q}}
\end{equation}
for $\tau$ positive \cite{bol:ssd}.  This is closely related to the form factor of the Laplace operator \cite{kot:qcg,kot:pot},
\begin{equation}\label{eq:KLaplace}
K_{\textrm{Laplace}}(\tau) = \frac{1}{(2L)^2}
\sum_{p,q} \frac{l_{p}l_{q}}{r_{p}r_{q}} A_{p} A_{q} \,
\ue^{\ui\pi(\mu_p +\mu_q)} \,
\delta \left( \tau  - \frac{l_{q}}{2L} \right) \delta_{l_{p},l_{q}} \ .
\end{equation}
Generically, for time-reversal symmetric boundary conditions, the form factor of the Dirac operator converges to the limiting distribution of the GSE when one considers the limit of a sequence of graphs with increasing numbers of edges.  In the same way the form factor of the Laplace operator on the graph converges to the GOE form factor.  Comparing power series expansions of the random matrix form factors for $\tau<1$ shows that they are also closely related \cite{heu:sol},
\begin{equation}
\begin{split}
K_{\GSE}(\tau) & =  \frac{\tau}{2} + \frac{\tau^2}{4} + \frac{\tau^3}{8}
+ \frac{\tau^4}{12} + \dots \ ,\\
\frac{1}{2} K_{\GOE}\left( \frac{\tau}{2} \right)
& =  \frac{\tau}{2} - \frac{\tau^2}{4} + \frac{\tau^3}{8}
- \frac{\tau^4}{12} + \dots \ .
\end{split}
\end{equation}
Calling $K^{m}$ the term containing $\tau^m$ the relationship can be
written
\begin{equation}\label{eq:RMT relation}
K^{m}_{\GSE}(\tau) = \left( -\frac{1}{2} \right)^{m+1} K^{m}_{\GOE}(\tau) \ .
\end{equation}

In (\ref{eq:KDirac}) and (\ref{eq:KLaplace}) a pair of orbits $p,q$ only contributes to the form factor if $l_p=l_q$.
Much of the success studying the form factor via periodic orbits has come by identifying sets of orbit pairs where the partner orbit of $p$ is generated by permuting the arcs of $p$ between self intersections and or reversing the direction of arcs.  Clearly permuting the order of edges of an orbit on a graph or reversing the direction a sequence of edges is traversed does not change the orbit length and pairs constructed in this manner must contribute to the form factor.  The first example of this approach in a chaotic quantum system was the diagonal approximation used by Berry \cite{ber:sts} to obtain the first term in the power series expansion of the form factor by evaluating the contributions of an orbit with itself, $p=q$ or with its time reversed partner using the sum rule of Hannay and Ozorio de Almeida \cite{han:poc}.
Sieber \cite{sie:lod} and Sieber and Richter \cite{sie:cbp} were the first to evaluate the second order contributions by considering orbits with one self-intersections.  This was extended to higher orders for quantum graphs in \cite{ber:lod,ber:fff} and recently for general contributions of orbit pairs with any number of self-intersections \cite{mul:pot}.

To evaluate such contributions to the form factor of a quantum graph requires us to consider the combined limit $L\to \infty$, $\tau \to 0$ with $L\tau \to \infty$ while the mean bond length $\overline{L}=L/N$ is kept constant.  In this limit long orbits dominate the sum and from these the proportion with $r_p\ne 1$ tends to zero and may safely be excluded.  We also approximate
$l_p\approx n\overline{L}$ where $n$ is the number of bonds in $p$.
 To isolate a spin contribution to the form factor
 (\ref{eq:KDirac}) we assign the spin transformations $u^{(v)}_{ef}$  randomly with Haar measure from a subgroup $\Gamma \subseteq \SU(2)$ so that the only correlation between the stability factor $A_p$ and the spin rotation $d_p$ comes from the common structure of the orbit.  Let $D$ be a set of pairs of orbits $(p,q)$ of length $n$ where each partner orbit $q$ is obtained from $p$ by the same permutation and or reversal of arcs of $p$.
 We can factor the sum over the orbit pairs in $D$ so,
 \begin{equation}
 \begin{split}
 \label{eq:factorK}
    \frac{1}{|D|} \sum_{D} A_{p} A_{q} \,
&\ue^{\ui\pi(\mu_p +\mu_q)} \tr (d_{p}) \tr (d_{q}) \\ & \sim
\left(  \frac{1}{|D|} \sum_{D} A_{p} A_{q} \,
\ue^{\ui\pi(\mu_p +\mu_q)} \right)
\left( \frac{1}{|D|} \sum_{D} \tr (d_{p}) \tr (d_{q}) \right) \ .
\end{split}
 \end{equation}
For long orbits and randomly chosen spin transformations we argue that
\begin{equation}\label{eq:spincontribution}
    \frac{1}{|D|} \sum_{D} \tr (d_{p}) \tr (d_{q})
    = \la \tr (d_{p}) \tr (d_{q}) \ra_\Gamma
\end{equation}
the average over $\Gamma$ with Haar measure.  This provides a sketch of how a spin contribution for certain sets of orbits can be isolated from terms in the form factor sum.  The remaining term in (\ref{eq:factorK}) is the contribution of these orbit pairs to the form factor of the Laplace operator on the graph.

In \cite{bol:scf,bol:sff} we calculate the spin contribution of such sets of orbit pairs.
\begin{theorem}
 Consider a pair of orbits $p,q$ on a graph where $q$ is related to $p$ by a permutation of arcs of $p$ and or reversing the direction of arcs.  Let the spin transformations on the graph generate a subgroup
 $\Gamma \subseteq \SU(2)$.
 \begin{equation*}\label{eq:SU(2)average}
    \la d_p , d_q \ra_\Gamma = \left( -\frac{1}{2} \right)^{m-1} \ ,
 \end{equation*}
 where the average is with respect to Haar measure.  The exponent $m-1$ is the number of self intersections at which the orbit $p$ has been rearranged to produce $q$.
\end{theorem}
Multiplying by the factor of $1/4$ from the form factor of the Dirac operator we obtain $\left( -\frac{1}{2} \right)^{m+1}$.  This generates the same
relationship as exists between the form factors of the GSE and GOE (\ref{eq:RMT relation}) provided that pairs of orbits where the partner $q$ differs from $p$ at $m-1$ self intersections contribute at order $\tau^m$ in a power series expansion of $K(\tau)$.  This is precisely the relationship found in derivations of the form factor without spin from periodic orbit theory
\cite{ber:lod,ber:fff,ber:sts,mul:pot,sie:lod,sie:cbp}.  The spectral statistics of quantum graphs with spin are investigated using a completely different technique in \cite{gnu:sci}.  Here the authors compute spectral correlations using super symmetric arguments and obtain the characteristic GSE statistics for a quantum graph with spin.  However, as the model of spin dynamics is specific it can't suggest conditions on the spin model  necessary for Random matrix statistics.

\section{Rashba localization}\label{sec:localization}

The effect of a spin orbit interaction can be found not only in spectral statistics but also in macroscopic properties of quantum
systems.  Spin orbit coupling in quantum graphs can be used to generate localization phenomena in periodic structures.  In \cite{ber:rei,ber:req} Bercioux et. al. demonstrate localization, zero conductance, numerically for a diamond chain lattice with equal edge lengths, figure \ref{fig:diamond chain}.
They consider the Rashba Hamiltonian on the graph where the chain is embedded in the $x$-$y$ plane and the magnetic field is perpendicular to the plane, as described in section \ref{sec:rashba}.  In the absence of a magnetic field
the Rashba coupling can be tuned to so that spin transformations along the graph edges conspire to produce cancelation in the wave functions at the vertices of valency four.  An electron traveling across one of the cells, from $v$ to $w$ in figure \ref{fig:diamond chain}, can take two possible paths.  Along an edge $e$ it acquires a spin rotation
%
\begin{equation}\label{eq:Rrotation}
    d_e=\exp \big[-\ui 2k_R L_e ( \vsig \times \vn ).\ve \big] \ .
\end{equation}
along with the dynamical phase $\ue^{\ui k L_e}$.  One may consider the spin transformation (\ref{eq:Rrotation}) as a non-Abelian phase which gives rise to destructive interference only for the special case of the square.  Although the effect only appears for this particular geometry the phenomena is seen numerically to be stable under small perturbations of the edge lengths about a mean where they are all equal.  Consequently it would be amenable to experimental investigation \cite{ber:rei}.

\begin{figure}[tb]
\begin{center}
\setlength{\unitlength}{4cm}
\begin{picture}(2,0.6)
\put(0,0){\includegraphics[width=8cm]{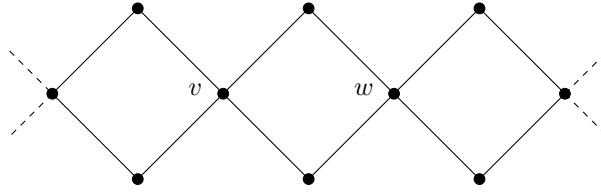}}
\put(0.6,0.3){$v$}
\put(1.15,0.3){$w$}
\end{picture}
\end{center}
\caption{A schematic view of the diamond chain.}
\label{fig:diamond chain}
\end{figure}

Figure \ref{fig:rashba} (a) shows the conductance across a diamond chain and a ladder as a function of the spin-orbit coupling averaged over the injection energies.  Localization in the form of destructive interference at the valency four vertices can also be generated by the Aharonov-Bohm phase in the absence of spin-orbit coupling  $k_R=0$ for particular values of the magnetic field strength.  These vertices which bound an initially localized wave-packet are known as an Aharonov-Bohm cage.  The same cage generated via the Rashba effect by tuning the spin-orbit coupling in the absence of a magnetic field.

\begin{figure}[tb]
\begin{center}
\setlength{\unitlength}{0.75cm}
\begin{picture}(13.5,4.3)
\put(0.5,0.3){\includegraphics[height=3cm]{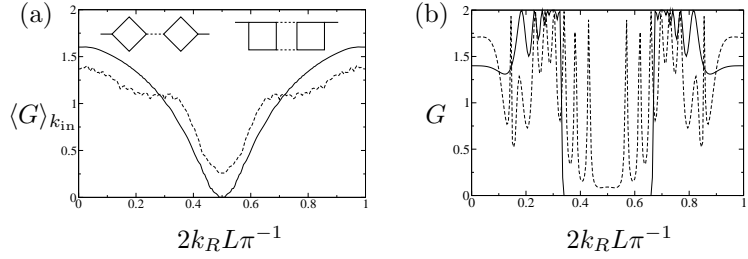}}
\put(0,4){(a)}
\put(2.8,0){$2k_R L\pi^{-1}$}
\put(-0.2,2.2){$\la G \ra_{k_{\mathrm{in}}}$}
\put(7.5,0.3){\includegraphics[height=3cm]{figure4.eps}}
\put(7,4){(b)}
\put(9.7,0){$2k_R L\pi^{-1}$}
\put(7.2,2.2){$G$}
\end{picture}	
\caption{(a) Conductance averaged over $k_{\mathrm{in}}$ as a function of the spin-orbit coupling strength for the diamond chain (continuous line) and for the ladder (dashed line).  The two finite-size systems connected to input/output leads are shown in the inset. The parameters used for the calculation are: 50 elementary loops, $k_{\mathrm{in}}$ uniformly distributed in $[0, \pi/L]$. (b) Conductance as a function of the spin-orbit coupling strength for the diamond chain (continuous line) and for the ladder (dashed line) for a fixed value of $k_{\mathrm{in}}=k_{\mathrm{F}}$. The parameters used for the calculation are: 50 elementary loops, $k_{\mathrm{F}} L=n \pi + 2$, being $n$ an integer.  Figure courtesy of Bercioux et. al. \cite{ber:rei}.\label{fig:rashba}}
\end{center}
\end{figure}

The spectrum and conductance of a Rashba Hamiltonian on the $\mathcal{T}_3$ lattice, see figure \ref{fig:t3}(a), was investigated numerically in \cite{ber:req}.  Spin-orbit coupling in this geometry no longer results in perfect localization but the Rashba effect can be tuned with or without the magnetic field to suppress the conductance, see figure \ref{fig:t3}(c).  Perfect localization is still possible when $k_R=0$, figure \ref{fig:t3}(b).

%
\begin{figure}[tb]
\begin{center}
\setlength{\unitlength}{0.72cm}
\begin{picture}(14,9.5)
\put(0.5,5.7){\includegraphics[width=4.32cm]{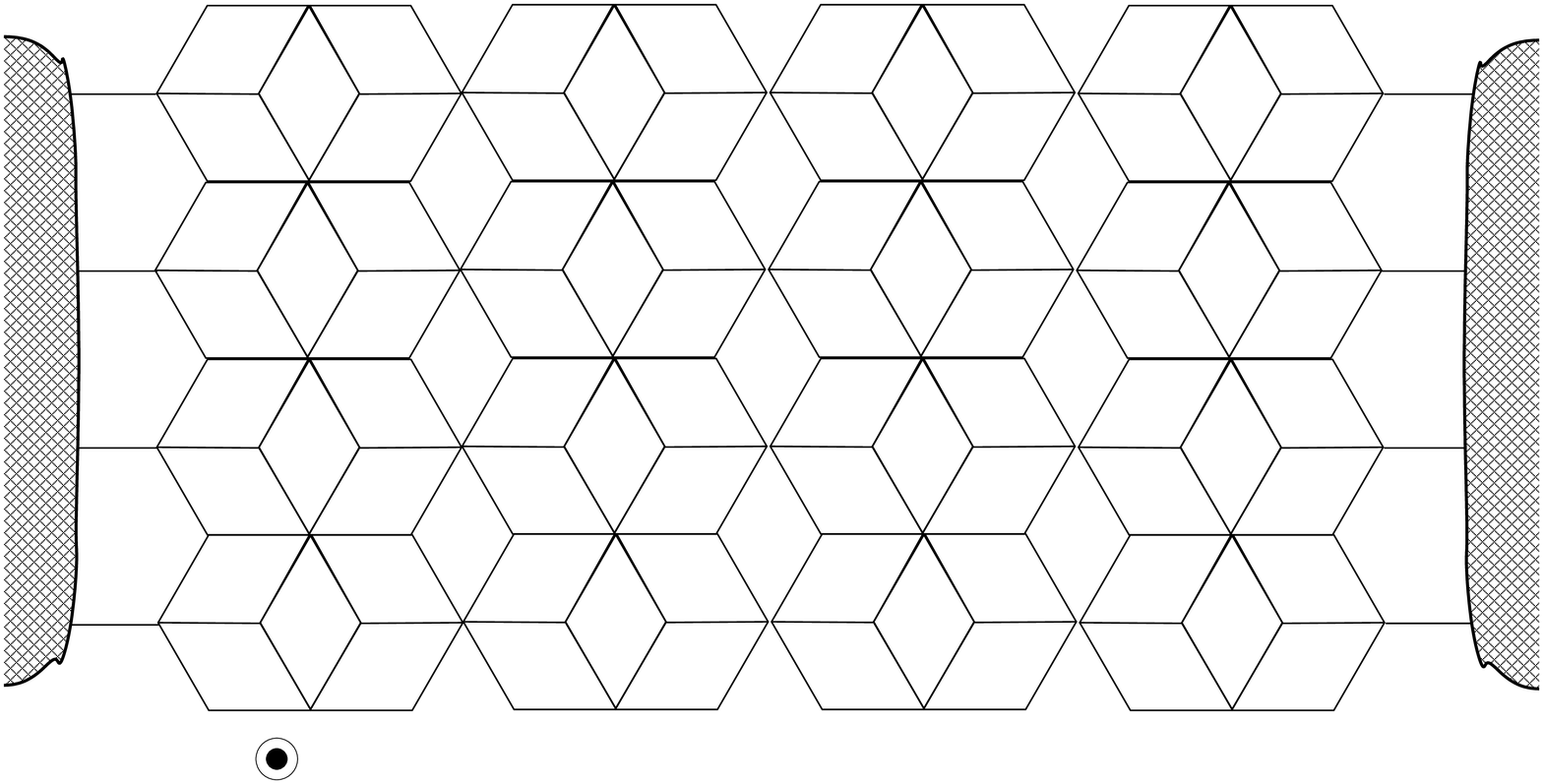}}
\put(0,9.5){(a)}
\put(1.8,5.5){$\mathbf{E},\mathbf{B}$}
\put(7.5,5.2){\includegraphics[width=4.32cm]{figure4b}}
\put(7,9.5){(b)}
\put(7,7.5){$\frac{\la G\ra_k}{N_\textrm{in}}$}
\put(7.8,9.5){{\small $2k_R L\pi^{-1}$ (spin-orbit coupling)}}
\put(9,5){{\small $\phi/\phi_0$ reduced flux}}
\put(0.5,0.35){\includegraphics[width=4.32cm]{figure4c}}
\put(0,4.5){(c)}
\put(0,2.5){$\frac{\la G\ra_k}{N_\textrm{in}}$}
\put(0.7,4.5){\small $2k_R L\pi^{-1}$  (spin-orbit coupling)}
\put(2,0){{\small $\phi/\phi_0$ reduced flux}}
\put(7.7,0.2){\includegraphics[width=4.32cm]{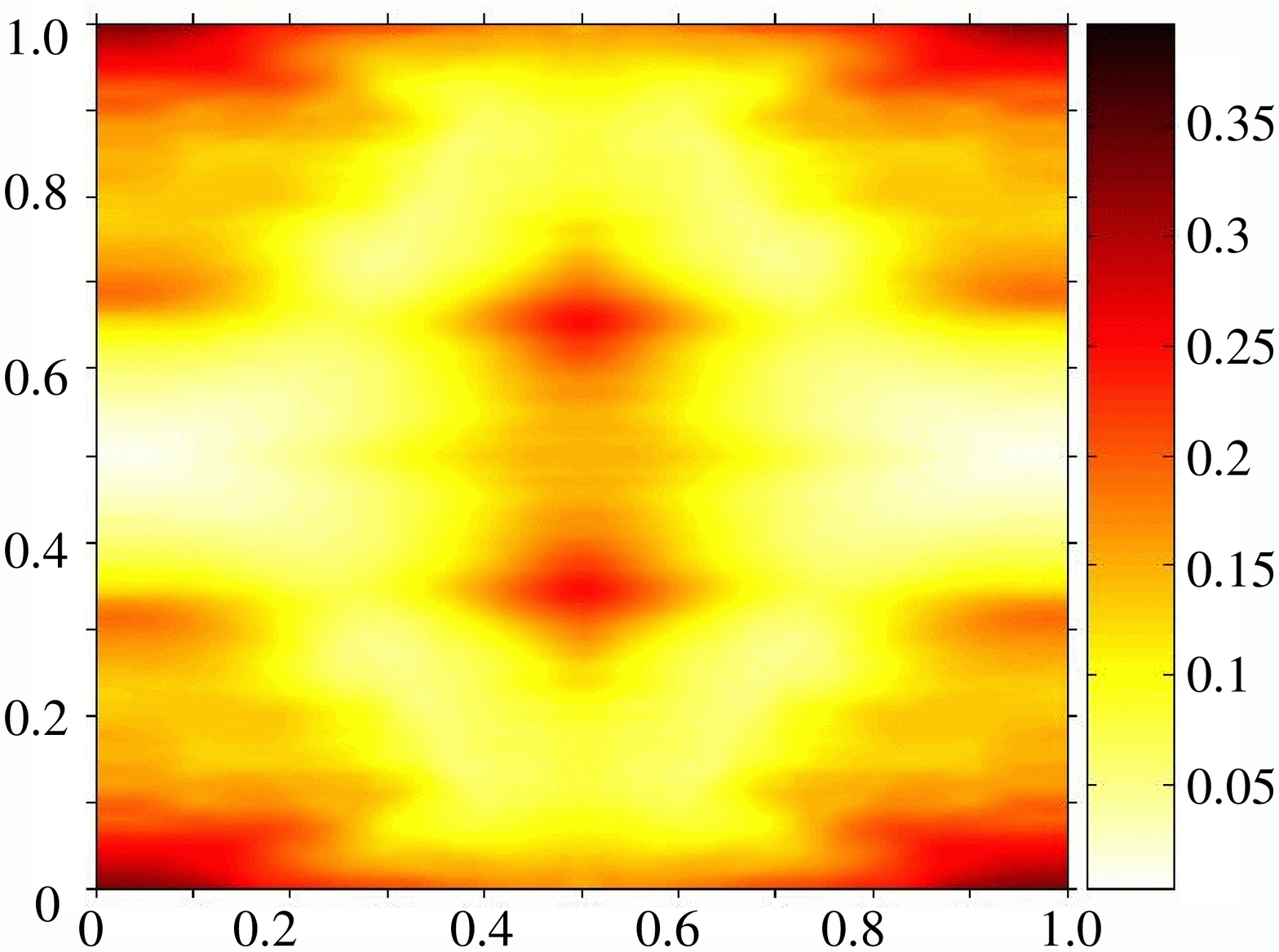}}
\put(7,4.5){(d)}
\put(7,2.5){{\small $\phi/\phi_0$}}
\put(10,0){{\small $2k_R L\pi^{-1}$}}
\end{picture}
\caption{(a): A finite-size piece of the $\mathcal{T}_3$ network
        connected to reservoirs.
        There are $4$ input and
        $4$ output single-channel leads.
        The number of input channels, considering spin, is $N_{\textrm{in}}=8$.
        (b): Averaged conductance
        per channel, $\langle G\rangle_k / N_{\textrm{in}}$, as a function of
        the reduced flux evaluated with $k_{R}= 0$
	(solid line), and of spin-orbit coupling evaluated with
$\phi/\phi_0 = 0$
        (dashed line) for the $\mathcal{T}_3$ lattice with $200$ quantum wires
        (89 rhombi). (c): Averaged conductance per
        channel, $\langle G\rangle_k / N_{\textrm{in}}$, as a function of the
        reduced flux evaluated at $2k_{R}L pi^{-1} = 0.5$
        (solid line), and of spin-orbit coupling evaluated at
        $\phi/\phi_0 = 0.5$ (dashed line).
        (d): Grey-scale plot of the
        averaged conductance per channel, $\langle G\rangle_k/N_{\textrm{in}}$, as a function
        of the reduced
        flux and spin-orbit coupling.  Figure courtesy of Bercioux et. al. \cite{ber:req}.\label{fig:t3}}
\end{center}
\end{figure}

In \cite{pan:lep} the spectrum of the infinite $\mathcal{T}_3$ lattice was characterized analytically via Bloch theory.
To give a flavor of the result we restrict to the case of the free Rashba Hamiltonian.  The matching conditions at vertices characterized in theorem \ref{thm:sa Rashba} are restricted to at most two types with a coupling constant $\epsilon(v)=\lambda$ at vertices of degree $6$ and $\epsilon(v)=\mu$ at vertices of degree $3$.
An interesting situation occurs when $\cos k_R=0$, i.e
$k_R\in \pi/2 +\pi \Z$ and the magnetic flux through a primitive rhombus $\omega= -\pi/6+\pi\Z$.
The spectrum may be characterized as the set of values of $E\in \R$ such that
\begin{equation*}\label{eq:T3bands}
\begin{split}
    \left( \cos \sqrt{E+k_R^2} +\frac{\lambda}{6\sqrt{E}} \sin \sqrt{E+k_R^2} \right)&.
    \left( \cos \sqrt{E+k_R^2} +\frac{\mu}{3\sqrt{E}} \sin \sqrt{E+k_R^2} \right)\\
   &\in \left[ 0, \frac{1}{6} \right]
   \cup \left\{ \frac{1}{3} \right\}
   \cup \left[ \frac{1}{2} , \frac{2}{3} \right] \ .
   \end{split}
\end{equation*}
along with solutions of the equation $\sin \sqrt{E+k_R^2}=0$.  In this situation together with the expected bands there appear infinitely degenerate eigenvalues.
Restricting further to Neumann like matching conditions, i.e. $\lambda=\mu=0$ the spectrum is the set of values of $E$ for which,
\begin{equation*}
    \cos^2 \sqrt{E+k_R^2} \in \left[ 0, \frac{1}{6} \right] \cup \left\{ \frac{1}{3} \right\} \cup \left[ \frac{1}{2}, \frac{2}{3} \right] \cup \{1\} \ .
\end{equation*}

Solutions of the equation
\begin{equation*}\label{eq:local}
    \left( \cos \sqrt{E+k_R^2} +\frac{\lambda}{6\sqrt{E}} \sin \sqrt{E+k_R^2} \right).
    \left( \cos \sqrt{E+k_R^2} +\frac{\mu}{3\sqrt{E}} \sin \sqrt{E+k_R^2} \right)=\frac{1}{3}
\end{equation*}
also appear as eigenvalues in the case of perfect localization induced by the magnetic field when $\omega \in \pi/2 +\pi\Z$ and $k_R\in \Z$.  Then all the bands collapse to a sequence of infinitely degenerate eigenvalues.
It is notable that by tackling localization problems on graphs the simplicity and symmetry of the problems can be exploited to achieve elegant analytic results.


\section*{Final remarks}\label{sec:con}
The article has been intended to survey techniques used to quantize graphs with spin.  Our motivation goes back to the origin of quantum graph models.  They were first proposed to describe electrons in organic molecules and modern experiments on mesoscopic networks also deal with the properties of electrons.  Our examples of applications of spin Hamiltonians show that the presence of a spin-orbit interaction on graphs has a significant influence on the models features.  A different class of spectral statistics or new modes of localization sensitive to the geometry of the network are both explained in terms of the spin rotations around periodic orbits.

To take graph models with spin further there remain important open questions.  Much has already been discovered about the convergence of Laplace operators on thin networks to the operator on a quantum graph and these results are sensitive to the boundary conditions at the surface of the domain.  Quantum graphs with spin-orbit interaction should provided more interesting models of mesoscopic systems.  To use them in this context it will be important to know when spin Hamiltonians converge to operators on quantum graphs.  It will also be interesting to discover the physically relevant matching conditions for the various spin Hamiltonians.
Then the ideal choice of model should introduce spin-orbit interactions in a way that maintains the simplicity that drew people to study graph models and provides realistic results that can guide future analysis and mesoscopic experiments.  So far the models generally fall into two classes: those where the spin transformation is at the vertices specified by matching conditions or those where spin rotation occurs along the edges.  It seems likely that these may be equivalent as the convergence results for operators on thin networks will likely produce matching conditions dependent on the graph geometry at the vertices.

\subsection*{Acknowledgements}

This work is supported by the National
Sciences Foundation under research grant DMS-0604859.
The survey was prompted by the authors visit to
the Isaac Newton Institute for Mathematical Sciences, Cambridge, UK,
which was partially supported by National Sciences
Foundation grant DMS-0648786.

\def\Dbar{\leavevmode\lower.6ex\hbox to 0pt{\hskip-.23ex \accent"16\hss}D}

\end{document}